\documentclass[12pt]{article}
\usepackage{a4}
\usepackage{epsfig,url}
\usepackage{lineno}

\newcommand\oo{\omega_{12}}
\newcommand\oon{\omega_{0}}
\newcommand\ooe{\omega_{E}}
\newcommand\Do{\Delta\omega}

\newcommand\dt{\Delta t}
\begin{document}
\title{A new method to measure the unloaded quality factor of superconducting cavities}
\author{Volker Ziemann, Uppsala University, 75120 Uppsala, Sweden}
\date{March 4, 2024}
\maketitle
\begin{abstract}\noindent
  We describe a method that measures the unloaded quality factor $Q_0$, the
  external quality factor $Q_E$, and the cavity detuning $\Do$ with a recursive
  least-squares algorithm. It combines a large number of consecutive
  measurements to successively improve an estimate of fit parameters that
  asymptotically converges to the ``real'' values. Exploiting the large amount
  of data acquired by a digital low-level radio frequency system permits us to reach this 
  asymptotic regime in a moderate time frame of seconds to minutes. Simulations
  show that the method works both for critically coupled and over-coupled
  cavities. A new calibration method addresses very tight tolerances of the
  method on system parameters.  
\end{abstract}
%
%
\section{Introduction}
The unloaded quality factor of superconducting cavities, commonly referred to as $Q_0$,
parameterizes the losses due to the energy dissipated in the walls of the cavity. These
dynamic losses constitute a major heat load for the cryogenic system, especially for
accelerators operating in continuous-wave mode~\cite{CEBAF,CBETA,LCLS2}, because it limits
their energy efficiency. Measuring $Q_0$ is therefore of paramount importance for the
successful operation of superconducting accelerating structures and the accelerators that
use them.
\par
Measuring $Q_0$ in critically coupled cavities is usually done in a vertical cryostat
and is based on comparing the incident power to the power escaping through the power
coupler, the measuring antenna, and via losses in the cavity walls~\cite{POWERS}. A
careful analysis of the associated uncertainties is documented in~\cite{MELNYCHUK}.
Those were found to be in the several-percent range. Additional systematic errors, especially
due to re-reflections from the circulator are discussed in~\cite{HOLZ2016,HOLZ2019}.
Refined methods, exploiting both amplitude and phase information, are discussed
in~\cite{FROLOV,GORYASHKO}.
\par
Measuring $Q_0$ in over-coupled cavities equipped with high-power couplers, usually
found in cryomodules, is more difficult. In particular, because the power escaping
through the power couplers is orders of magnitude larger than the power dissipated in
the cavity walls or escaping through the measuring antenna. In the past this made it practically
impossible to measure $Q_0$  by analyzing radio-frequency (RF) signals, where~\cite{KUZIKOV}
is an exception. Therefore, calorimetric measurements are normally used. Either by measuring
the increase of the helium pressure~\cite{DRURY}, if the enclosing helium vessel is closed,
or by measuring the flow of evaporated helium~\cite{WANG}, if helium cooling is operated
in steady state. The situations is aggravated, if multiple cavities are jointly connected
to a single helium vessel, which only allows to measure their combined power dissipation.
The cited accuracies for the determination of $Q_0$ are in the 10\,\% to
15\,\% level, reached after averaging the helium flow for times measured in hours.
\par
In this report, we extend the analysis from~\cite{SYSID} and describe a new method that
measures the unloaded quality factor $Q_0$, the external quality factor $Q_E$, and the
cavity detuning $\Do$ for critically coupled and for over-coupled cavities. It uses
a recursive least-squares algorithm and combines a large number of successive measurements
to improve an estimate of the fit parameters. Affected by process noise only, this type
of measurement can be shown to asymptotically converge to the ``real'' values~\cite{LAIWEI},
whereas measurement noise can slightly bias the asymptotic value. We will discuss this further
below. Nevertheless, exploiting the large amount of data acquired by a digital low-level RF
system permits us to reach the asymptotic regime in a moderate time frame of seconds or minutes.
\par
This report is organized as follows. We first discuss the model and the system identification
process in Section~\ref{sec:sysid}, before simulating a generic system in Section~\ref{sec:sim},
where we find that, especially for over-coupled systems, the sensitivity to systematic errors
is extreme. We address this problem in Section~\ref{sec:cal} with a new calibration method,
specific for this type of measurement, before concluding.
\section{Model and system identification}
\label{sec:sysid}
We describe the superconducting cavity as resonator with resistor $R$, capacitor $C$,
and inductor $L$ coupled in parallel that is driven by a current $I$. After averaging
over the RF oscillations, the envelope of the voltage across the components is described
by $V=V_r + iV_i$, where $V_r$ is the real part of the voltage (in-phase, I) and $V_i$
is the imaginary part (quadrature phase, Q). This dynamical system can be written
as~\cite{SCHILCHER}
\begin{equation}\label{eq:sss}
  \left(\begin{array}{c} \frac{dV_r}{dt}\\ \frac{dV_i}{dt} \end{array}\right)
  =\left(\begin{array}{cc} -\oo & -\Do \\ \Do& -\oo \end{array}\right)
  \left(\begin{array}{c} V_r \\ V_i \end{array}\right)
  +\left(\begin{array}{cc} \oo R & 0 \\ 0& \oo R \end{array}\right) 
  \left(\begin{array}{c} I_r \\ I_i \end{array}\right)\ ,
\end{equation}
where $\Do$ is the cavity detuning, $\hat\omega$ is the resonance frequency of the
cavity, and $\oo=\hat\omega/2 Q_L$ is the cavity bandwidth with the loaded quality
factor $1/Q_L=1/Q_E+1/Q_0+1/Q_t$. Here $Q_E$ is the external quality factor, $Q_0$
the unloaded quality factor, and $Q_t\gg Q_0$ is the quality factor of the antenna
that measures the field in the cavity. $I=I_r+iI_i$ is the current that drives the
cavity. It is measured by directional couplers that determine the forward current
$I^+=I^+_r+iI^+_i$ and the reflected current $I^-=I^-_r+iI^-_i$. Close to resonance
$\vec I^+=(I^+_r,I^+_i)^{\top}$ is related to $\vec I=(I_r,I_i)^{\top}$ through
$\vec I = (2Q_L/Q_E)\vec I^+$, which leads us to~\cite{SYSID}
\begin{equation}\label{eq:ss}
  \left(\begin{array}{c} \frac{dV_r}{dt}\\ \frac{dV_i}{dt} \end{array}\right)
  =\left(\begin{array}{cc} -\oo & -\Do \\ \Do& -\oo \end{array}\right)
  \left(\begin{array}{c} V_r \\ V_i \end{array}\right)
  +\left(\begin{array}{cc} \ooe R & 0 \\ 0& \ooe R \end{array}\right) 
  \left(\begin{array}{c} I^+_r \\ I^+_i \end{array}\right)\ .
\end{equation}
Converting the system to discrete time with time step $\dt$ results in
\begin{equation}\label{eq:dsa}
  \vec V_{t+1} 
  =A \vec V_t  + B \vec I^+_t +\vec w_t
  \quad\mathrm{with}\quad A=\left(\begin{array}{cc} 1-\oo \dt & -\Do \dt\\ \Do \dt& 1-\oo \dt \end{array}\right)
\end{equation}
and $B=\ooe \dt R\mathbf{1}$. Here $\vec w_t$ is the uncorrelated process noise with magnitude $\sigma_p$. 
It is thus characterized by its expectation value $E\left\{\vec w_t\vec w_s^{\top}\right\}=\sigma_p^2\delta_{ts}\mathbf{1}$. 
Additionally, we add measurement noise $\vec w'_t$ by using
\begin{equation}\label{eq:mnoise}
  \vec V'_t=\vec V_t + \vec w'_t
\end{equation}
in the system identification process. We assume it is uncorrelated, has magnitude $\sigma_m$, 
and is characterized by $E\left\{\vec w'_t\vec w_s^{\prime\top}\right\}=\sigma_m^2\delta_{ts}\mathbf{1}$.
\par
Similar to~\cite{SYSID} we isolate the dependence on the fit parameters by first writing
Equation~\ref{eq:dsa} as
\begin{equation}\label{eq:yy}
  \vec V'_{t+1}-\vec V'_t  = F \vec V'_t+ \ooe\dt R\vec I^+_t
\end{equation}
with
\begin{equation}
  F=\left(\begin{array}{cc}
      -\frac{1}{2}(\ooe+\oon)\dt & -\Do\dt \\ \Do\dt & -\frac{1}{2}(\ooe+\oon)\dt       
    \end{array}\right)
\end{equation}
and $\oon=\hat\omega/Q'_0$ with $1/Q'_0=(1/Q_0+1/Q_t)$. That $\omega_0$ depends on both
$Q_0$ and $Q_t$ indicates that we cannot disentangle their respective contributions using
this method. Since normally $Q_t\gg Q_0$ this should not be a big problem. Otherwise,
$\beta_t=Q_0/Q_t$ must be determined by conventional methods~\cite{POWERS,MELNYCHUK}
of the critically coupled cavity. Continuing with the discussion of the system
identification process, we write the right-hand side of Equation~\ref{eq:yy} as
\begin{eqnarray}
 F\vec V'_t + \ooe\dt R\vec I^+_t
 &=& \ooe\dt \left(\begin{array}{c} -\frac{1}{2}V'_r+R I^+_r \\ -\frac{1}{2}V'_i+R I^+_i \end{array}\right)_t\nonumber\\
&&\qquad
  + \Do \dt \left(\begin{array}{r} -V'_i \\ V'_r \end{array}\right)_t
  +\oon\dt \left(\begin{array}{c} -\frac{1}{2}V'_r \\ -\frac{1}{2}V'_i \end{array}\right)_t\\
  &=& \left(\begin{array}{rrr} 
     -\frac{1}{2}V'_r+R I^+_r & -V'_i & -\frac{1}{2}V'_r  \\ 
     -\frac{1}{2}V'_i+R I^+_i & V'_r & -\frac{1}{2}V'_i\end{array}\right)_t
  \left(\begin{array}{c} \ooe \dt \\ \Do \dt \\ \oon\dt \end{array}\right)\ .\nonumber
\end{eqnarray} 
This allows us to write Equation~\ref{eq:yy} in the compact form
\begin{equation}\label{eq:yy2}
\vec y_{t+1} = G_t \left(\begin{array}{c} \ooe \dt \\ \Do \dt \\ \oon\dt \end{array}\right)
\end{equation}
with 
\begin{equation}\label{eq:defG}
G_t = \left(\begin{array}{rrr} 
     -\frac{1}{2}V'_r+R I^+_r & -V'_i & -\frac{1}{2}V'_r  \\ 
     -\frac{1}{2}V'_i+R I^+_i & V'_r & -\frac{1}{2}V'_i\end{array}\right)_t
\qquad\mathrm{and}\qquad 
\vec y_{t+1}=  \vec V'_{t+1}-\vec V'_t 
\end{equation}
which has the same structure already encountered in~\cite{SYSID}. Following in
the same vein we write successive versions of Equation~\ref{eq:yy2} in a vectorized 
form 
\begin{equation}\label{eq:UT}
  \left(\begin{array}{c} \vec y_2 \\ \vec y_3 \\ \vdots \\ \vec y_{T+1}\end{array}\right)
  = U_T  \left(\begin{array}{c} \ooe \dt \\ \Do \dt \\ \oon\dt  \end{array}\right)
  \quad\mathrm{with}\quad
  U_T=\left(\begin{array}{c} G_1 \\ G_2 \\ \vdots \\ G_T\end{array}\right)
\end{equation}
and solve it with the Moore-Penrose pseudoinverse~\cite{PENROSE} to arrive at
\begin{equation}\label{eq:ls}
  \vec q_T=\left(\begin{array}{c} \ooe \dt \\ \Do \dt \\ \oon\dt \end{array}\right)_T=
  \left( U_T^{\top} U_T\right)^{-1} U_T^{\top}
  \left(\begin{array}{c} \vec y_2 \\ \vec y_3 \\ \vdots \\ \vec y_{T+1}\end{array}\right) \ ,
\end{equation}
where we introduce the vector $\vec q_T$ of the estimate of fit parameters after $T$
iterations.
\par
The matrix $U_T$ rapidly grows to unmanageable proportions such that we resort to 
a recursive algorithm that solves it one step at a time. We therefore introduce 
$P_T^{-1}=U^{\top}_TU_T$ and its initial value $P_0=p_0{\mathbf 1}$ and express $P_{T+1}$ 
through $P_T$ in the following way
\begin{eqnarray}\label{eq:PTT}
  P_{T+1}^{-1}&=&U^{\top}_{T+1}U_{T+1}\nonumber\\
              &=&p_0{\mathbf 1}+G_1^{\top}G_1+G_2^{\top}G_2+\dots+G_T^{\top}G_T+G_{T+1}^{\top}G_{T+1}\\
              &=& P_T^{-1}+G_{T+1}^{\top}G_{T+1}\ . \nonumber
\end{eqnarray}
And here the previously mentioned bias originates.
Since the matrices $G_t$ depend linearly on the voltages $\vec V'_t$, the terms $G^{\top}_tG_t$
depend quadratically on $\vec V'_t=\vec V_t+\vec w'_t$. The measurement noise $\vec  w'_t$
from Equation~\ref{eq:mnoise} therefore also appears quadratically and on average causes
a bias proportional to $\sigma_m^2$, which turned out to be insignificant in all simulations
reported below.
\par
It turns out that retrieving $P_{T+1}$ from inverting $P_T^{-1}+G_{T+1}^{\top}G_{T+1}$ can be
efficiently done with the Woodbury matrix identity~\cite{NR}
\begin{equation}
  (A+VW^{\top})^{-1}=A^{-1}- A^{-1}V(\mathbf{1}+W^{\top}A^{-1}V)^{-1}W^{\top} A^{-1}\ .
\end{equation}
With the substitutions $A^{-1}=P_T$, $V=G^{\top}_{T+1}$, and $W^{\top}=G_{T+1}$, we obtain
\begin{equation}\label{eq:upp}
  P_{T+1}= \left[\mathbf{1}-P_TG_{T+1}^{\top}\left(\mathbf{1}+G_{T+1}P_TG_{T+1}^{\top}\right)^{-1} G_{T+1}\right]P_T\ .
\end{equation}
We now continue to find $\vec q_{T+1}$ by writing Equation~\ref{eq:ls} for $T+1$
\begin{eqnarray}\label{eq:upq}
  \vec q_{T+1}&=& P_{T+1} \left(G_1^{\top}\vec y_2+G_2^{\top}\vec y_3+\dots
                  + G_T^{\top}\vec y_{T+1} +G_{T+1}^{\top}\vec y_{T+2}\right)\nonumber\\
              &=&\left[\mathbf{1}-P_TG_{T+1}^{\top}\left(\mathbf{1}+G_{T+1}P_TG_{T+1}^{\top}\right)^{-1} G_{T+1}\right]P_T\\
                  &&\qquad\times
                  \left(\sum_{t=1}^T G_t^{\top}\vec y_{t+1} +G_{T+1}^{\top}\vec y_{T+2}\right)\nonumber\\
              &=&\left[\mathbf{1}-P_TG_{T+1}^{\top}\left(\mathbf{1}+G_{T+1}P_TG_{T+1}^{\top}\right)^{-1} G_{T+1}\right]
                  \left( \vec q_T+P_tG_{T+1}^{\top}\vec y_{T+2}\right)\ . \nonumber
\end{eqnarray}
Equation~\ref{eq:upp} and~\ref{eq:upq} constitute the algorithm
to continuously update estimates for the three components of $\vec q$, the bandwidth
$q(1)=\ooe \dt$, the detuning $q(2)=\Do \dt$, and $q(3)=\oon\dt$ as new voltage and current 
measurements become available.
\par
Like in~\cite{SYSID} we introduce a ``forgetting factor'' $\alpha=1-1/N_f$, where $N_f$ is
the time horizon over which old measurements are discounted. This is of limited practical
value here, because it limits the achievable accuracy of the method. It might, however,
become useful to investigate the stability of the
measurements and whether some unaccounted external factors cause the fit parameters to
vary over time. To avoid the limitation of the accuracy in this report, we choose $N_f=10^6$
in all simulations, a value much larger than the number of iterations used.
With $\alpha$ included, Equation~\ref{eq:PTT} reads $P_{T+1}^{-1}= \alpha P_T^{-1}+G_{T+1}^{\top}G_{T+1}$
and Equation~\ref{eq:upp} becomes
\begin{equation}\label{eq:uppt}
  P_{T+1}= \frac{1}{\alpha}\left[\mathbf{1}-P_TG_{T+1}^{\top}
          \left(\alpha+G_{T+1}P_TG_{T+1}^{\top}\right)^{-1} G_{T+1}\right]P_T
\end{equation}
whereas Equation~\ref{eq:upq} turns into
\begin{equation}\label{eq:upqt}
  \vec q_{T+1}=\left[\mathbf{1}-P_TG_{T+1}^{\top}\left(\alpha\mathbf{1}+G_{T+1}P_TG_{T+1}^{\top}\right)^{-1} G_{T+1}\right]
                  \left( \vec q_T+\frac{1}{\alpha}P_tG_{T+1}^{\top}\vec y_{T+2}\right)\ .
\end{equation}
In these equations the calculations are moderately time-consuming, because the involved quantities
are matrices. In particular, $P_T$ is a $3\times 3$ matrix, $G_{T+1}$ is a $2\times 3$ matrix, and 
the inversion involves the $2\times 2$ matrix $\alpha\mathbf{1}+G_{T+1}P_TG_{T+1}^{\top}$. 
\section{Simulations}
\label{sec:sim}
In the simulations we assume a generic superconducting cavity with a resonance frequency
$\hat\omega/2\pi=10^9$\,Hz and $Q'_0=10^9$. As in~\cite{SYSID} we scale the voltages and currents to their 
maximum values and denote the corresponding real and imaginary values by $v_r$, $v_i$, 
$i_r$, and $i_i$ respectively. For the process noise we assume $\sigma_p=10^{-4}\times V_{max}$
and for the measurement noise $\sigma_m=10^{-3}\times V_{max}$.
Once the cavity reaches a steady state, the first 
and last column of $G_t$ in Equation~\ref{eq:defG} become linearly dependent. We therefore
have to pulse the generator that excites the cavity. In all simulations we turn it on and 
off every 1000 iterations. We refer to the MATLAB~\cite{MATLAB} code on github~\cite{GITHUB}
for the details of the implementation.
\par
\begin{figure}[tb]
\begin{center}
  \includegraphics[width=0.47\textwidth]{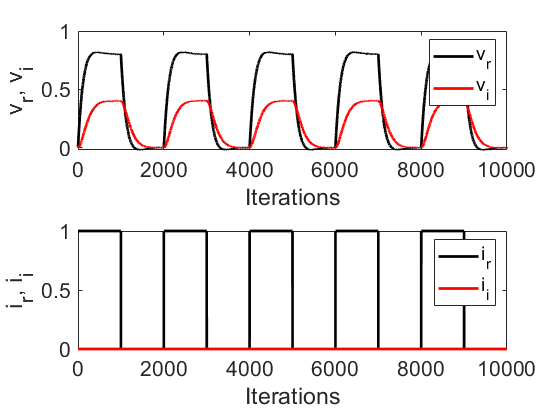}
  \includegraphics[width=0.47\textwidth]{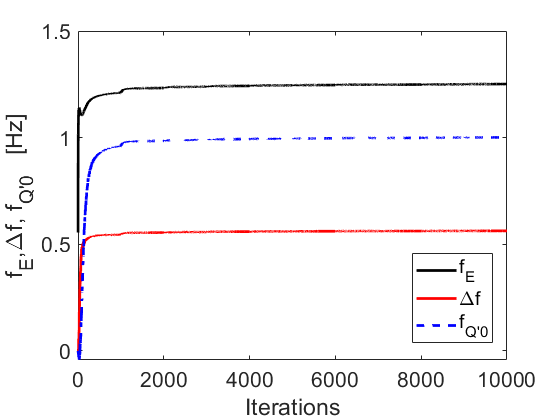}
  \includegraphics[width=0.47\textwidth]{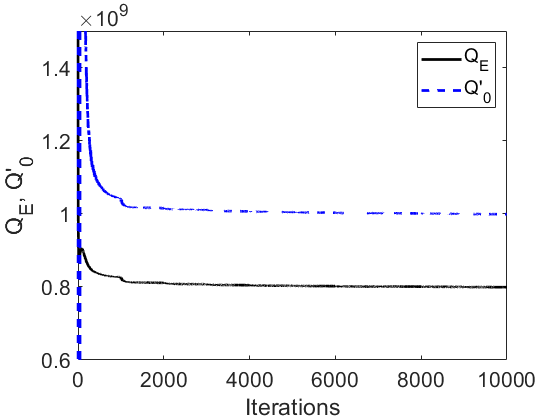}
  \includegraphics[width=0.47\textwidth]{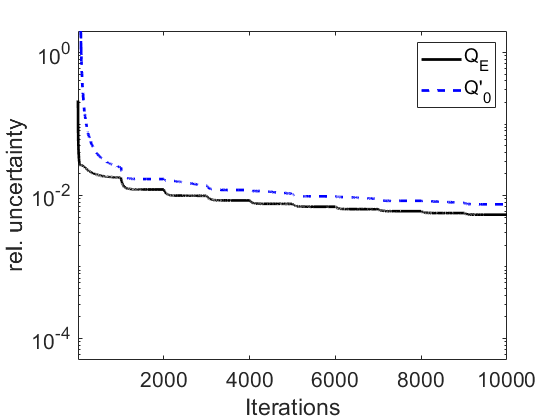}
\end{center}
\caption{\label{fig:crit}Simulation of a critically coupled cavity: voltages
  and currents (top left); fit parameters $f_E=\ooe/2\pi$, $\Delta f=\Do/2\pi$,
  and $f_{Q_0}=\oon/2\pi$ (top right); $Q_E$ and $Q'_0$ (bottom left); relative
  uncertainties (bottom right).}
\end{figure}
We first consider a close-to-critically-coupled system and assume $Q_E=8\times 10^8$, which results 
in a bandwidth of $f_{12}=\omega_{12}/2\pi=1.125\,$Hz. For the detuning we assume $\Do=\oo/2$.
Since both $\ooe$ and $\oon$ are very small we chose a moderately slow sampling rate of
1\,kSample/s or $\dt=10^{-3}\,$s. If the LLRF system operates at a higher rate, low-pass
filtering and subsequent undersampling the output is beneficial because it reduces the
measurement noise and reduces the bias mentioned before.
\par
The top-left part in Figure~\ref{fig:crit} shows the normalized voltages and currents
for 10000 iterations, which corresponds to 10\,s in real time. The pulsing currents
(lower panel) and the responding voltages (upper panel) are clearly visible. The top-right plot
shows $f_E=\ooe/2\pi$, $\Delta f=\Do/2\pi$, and $f_{Q_0}=\oon/2\pi$ as the simulation progresses.
We observe that these fit parameters approach constant values after a few thousand iterations.
The lower-left part shows $Q_E=\hat\omega/\ooe$ and $Q'_0=\hat\omega/\oon$. Both quantities
approach the correct values on the same time scale as the frequencies. The part on the
lower right shows the relative uncertainties $\sigma(Q)/Q$ (the ``relative error bars'')
derived from the diagonal elements of the empirical covariance matrix $P_T$. For example
the error bar of $Q'_0$ is approximately given by $\sigma(Q'_0)=\sqrt{P_T(3,3)}\sigma_m$.
We find that these statistical uncertainties quickly approach the percent level.
\par
We evaluate the sensitivity to systematic uncertainties by either changing the resistance
$R$ or the calibration scale factor to determine $\vec V'_T$ by 5\,\% when producing the data, thus
causing a systematic mismatch between the assumed and the ``real'' model. We find
that these changes also affect the fitted values of $Q_E$ and $Q'_0$ by about 5\,\%.
We conclude that the method is reasonably robust in this regime of operation.
\par
\begin{figure}[tb]
\begin{center}
  \includegraphics[width=0.47\textwidth]{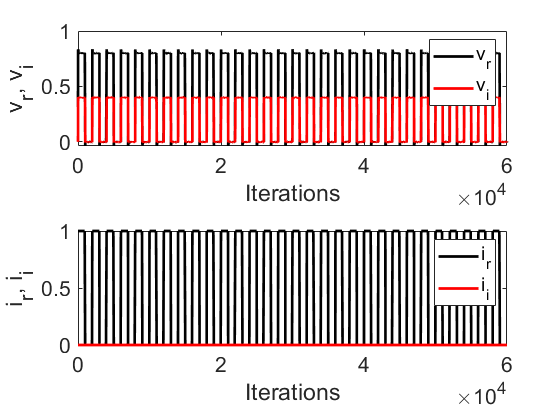}
  \includegraphics[width=0.47\textwidth]{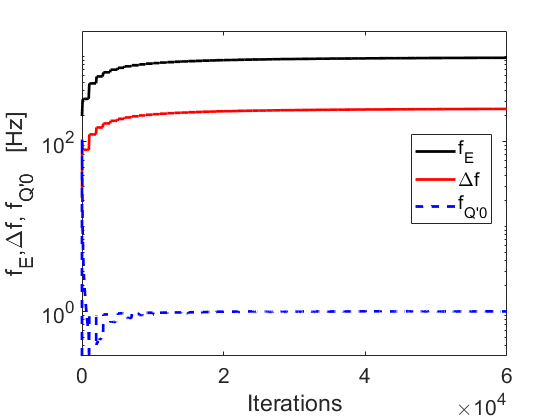}
  \includegraphics[width=0.47\textwidth]{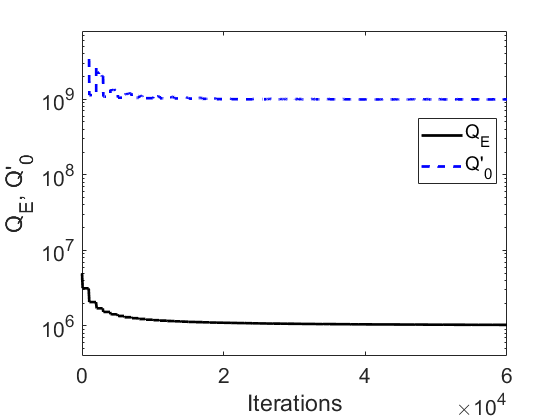}
  \includegraphics[width=0.47\textwidth]{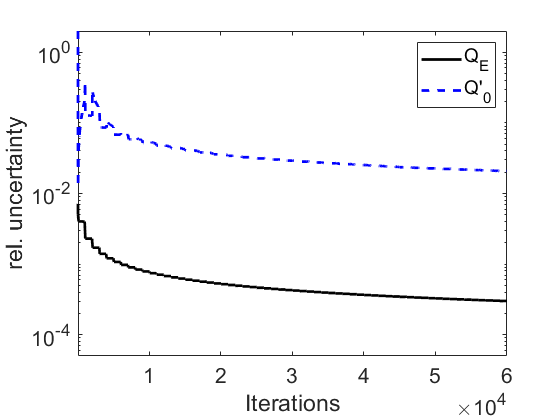}
\end{center}
\caption{\label{fig:over}Simulation of an over-coupled cavity: voltages
  and currents (top left); fit parameters (top right); $Q_E$ and $Q'_0$ (bottom
  left); relative uncertainties (bottom right).}
\end{figure}
Next we consider an over-coupled cavity with $Q_E=10^6$,  which results in a bandwidth
of $f_{12}=\omega_{12}/2\pi=500\,$Hz. Again we use $\Do=\oo/2$ in the simulations. 
Since $Q_E$ is much lower, the system reacts much more speedily and we choose a smaller
value of $\dt=10^{-4}\,$s. The simulation extends over $6\times 10^4$ iterations which
corresponds to six seconds real time.
\par
The four parts in Figure~\ref{fig:over} follow the same order as in Figure~\ref{fig:crit}
where the top left shows the pulsing voltages and currents and the top right shows the
fit parameters with $\oon/2\pi\approx 1\,$Hz and $\ooe/2\pi \approx 1000\,$Hz due to the
large difference between $Q_E$ and $Q'_0$. These frequencies lead to the fit values for
$Q'_0$ and $Q_E$ shown on the bottom left. Within about $2\times 10^4$ iterations they
approach the values used to generate the data. The corresponding relative uncertainties,
shown on the bottom right, are 2\,\% and less than $10^{-3}$, respectively. We conclude
that the algorithm actually recovers the quality factors with good resolution, despite
their three-order-of-magnitude difference in magnitude.
\par
The sensitivity to calibration uncertainties of the resistance $R$ and voltage
measurement, on the other hand, is extreme. Scale errors, off by $5\times 10^{-4}$,
either of the shunt impedance or the voltage measurements scale almost double the fitted value
of $Q'_0$. Repeating the sensitivity analysis for a cavity with $Q_E=10^7$ we find that
ten times larger scale errors of $5\times 10^{-3}$ double the fitted value of $Q'_0$,
whereas in a cavity with $Q_E=10^8$ scale errors of $5\times 10^{-2}$ double $Q'_0$.
\par
This extreme sensitivity to systematic errors warrants a discussion of methods
to calibrate the hardware.
\section{Calibration}
\label{sec:cal}
We emphasize that all hardware must be meticulously calibrated following the
discussion in~\cite{POWERS,MELNYCHUK} as well as calibrating the propagation
delays by de-embedding~\cite{AN1364-1} the ancillary hardware.

Once this is accomplished we turn our attention to the new algorithm and note
that it depends entirely on the definition of the matrix $G_t$ from
Equation~\ref{eq:defG}. We observe that any
scale factors of currents or voltages can be absorbed in the resistance $R$, which
serves as a factor to make currents and voltages commensurate. The systematic errors
discussed at the end of the previous section were caused by such an imbalance between
the currents and voltages that we can compensate by adjusting $R$. For example,
simultaneously increasing all voltages ten-fold and, at the same time, increasing
$R$ ten-fold, leaves the fit-parameters unaffected. This is easy to see, because
both $G_t$ and $\vec y_{t+1}$ from Equation~\ref{eq:defG} are multiplied by ten, but
this common factor cancels, once $\vec q_T$ is calculated in Equation~\ref{eq:ls}.
Being the only free parameter in the algorithm, we therefore have to determine $R$
to the $10^{-4}$ level.
\par
\begin{figure}[tb]
\begin{center}
  \includegraphics[width=0.47\textwidth]{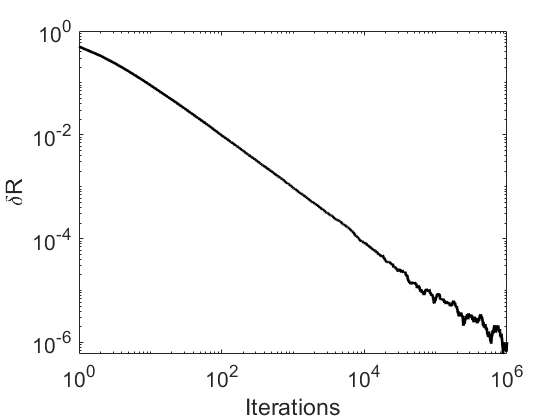}
  \includegraphics[width=0.47\textwidth]{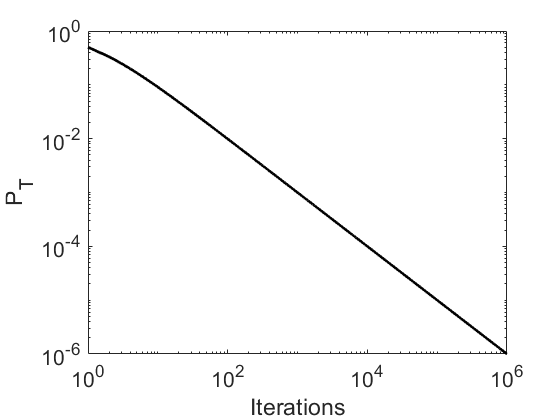}
\end{center}
\caption{\label{fig:cal}The estimation error (left) during the calibration of $R$
  and the empirical covariance matrix $P_T$ (right) as a function of the number of
  iterations.}
\end{figure}
We determine $R$ from Equation~\ref{eq:sss} in steady state ($dV/dt=0$) and
on-resonance ($\Do=0$). With input currents at a constant value, the steady state
is typically reached  after a few seconds. If the data acquisition system runs
with a high sampling rate, the algorithm from~\cite{SYSID} can validate that
$\Do$ is indeed zero. Under these conditions  Equation~\ref{eq:sss} simplifies to
\begin{equation}\label{eq:onres}
 0 =\left(\begin{array}{cc} -\oo & 0 \\ 0 & -\oo \end{array}\right)
  \left(\begin{array}{c} V_r \\ V_i \end{array}\right)_t
  +\left(\begin{array}{cc} \oo R & 0 \\ 0& \oo R \end{array}\right) 
  \left(\begin{array}{c} I_r \\ I_i \end{array}\right)_t\ .
\end{equation}
Note that here we use the total current $\vec I=\vec I^++\vec I^-$ instead of
the forward current $\vec I^+$. We thus need to add both channels from the
directional coupler normally used to measure the forward current only. By
inspecting Equation~\ref{eq:onres} we find that the bandwidth $\oo$ cancels,
leaving us with
\begin{equation}\label{eq:fitR}
  \left(\begin{array}{c} V'_r \\ V'_i \end{array}\right)_t
  =R\left(\begin{array}{c} I_r \\ I_i \end{array}\right)_t \ .
\end{equation}
In the latter equation we added the prime to the voltages to indicate that their
measurement is affected by measurement noise of magnitude $\sigma_m$. Moreover,
we explicitly added the subscript $t$ to denote that even in steady state the
noise causes the values to vary with time. Thus we can also repeat measurements
to improve the accuracy of determining $R$.
Equation~\ref{eq:fitR} has indeed the same structure as Equation~\ref{eq:yy2}
with $\vec V'_t$ taking the role of $\vec y_{t+1}$, the resistance $R$ taking the
role of $\vec q_T$, and $\vec I_t$ taking the role of $G_t$. We can therefore
use the same recursive least-squares algorithm already described in Section~\ref{sec:sysid},
especially Equations~\ref{eq:uppt} and~\ref{eq:upqt}, to determine $R$. The
corresponding equations now read
\begin{equation}
  P_{T+1}=\left[\frac{1}{\alpha + P_T \vec I_T^2}\right] P_T
\end{equation}
and
\begin{equation}
  q_{T+1} =\left[\frac{1}{\alpha + P_T \vec I_T^2}\right]
  \left(\alpha q_T + P_T \vec I^{\top}_T\vec V'_T\right)
\end{equation}
where $q_T$ is the steadily improving estimate of the resistance $R$. Again, we refer
to~\cite{GITHUB} for the details of the implementation in MATLAB.
\par
The left-hand plot in Figure~\ref{fig:cal} shows the difference between the estimated
value of $R$ and the true value, often referred to as estimation error, as the
calibration progresses for $10^6$ iterations after steady state was reached. Here we
do not use any forgetting factor and set $\alpha=1$. We observe that the estimation
error steadily decreases below the $10^{-4}$ level that is required to determine the
unloaded quality factor in over-coupled cavities. The right-hand plot shows the
empirical covariance matrix $P_T$ as the calibration progresses. It steadily
decreases as well. From $P_T$ we can obtain the approximate error bars of $R$ by
evaluating $\sqrt{P_T}\sigma_m$. 
\par
We thus find that the described calibration procedure indeed allows us to determine
$R$ with high precision. All systematic scale factors are then incorporated in $R$
as long as the hardware used in the calibration is the same one used to measure $Q'_0$.
\section{Conclusions}
\label{sec:conc}
We described a new method to determine the unloaded quality factor of superconducting
cavities. The method works both for critically coupled and over-coupled cavities, but
is very sensitive on systematic calibration errors. This sensitivity can be absorbed
into the resistance $R$ that can be calibrated with high precision. Both the
identification of the quality factors and the calibration of $R$ employ recursive
least-squares algorithms, which are known to converge towards the ``real'' values.
\par
We have, however, to keep in mind that the method relies on the fit parameters being
constant while the measurement is ongoing. This can be verified, provided the data
acquisition system and the analog-to-digital converters sample at a high rate. Low-pass
filtered and undersampled signals are passed to the algorithm described in this paper, while
the high-speed signals are passed to the algorithm described in~\cite{SYSID}. The latter
can then be operated in parallel to determine the detuning $\Do$ and the bandwidth
$\oo$ at high speed to ensure that the parameters are indeed constant. A rapid
increase in $\oo$ would indicate a quench, while microphonics or Lorentz force
detuning show up as variations of $\Do$.
\par
In particular at very high power levels, the Lorentz force detuning
could become a problem, especially due to pulsing the input current. This might limit
the new method to lower power levels. Optionally, other, more gentle excitation patterns,
for example by amplitude-modulating the input power with a sinusoidal variation, can
be explored. This would still ensure the algorithm to work, though possibly at the
expense of slower convergence. Optimizing these methods is, however, outside the scope
of this report.
\par
The need to pulse or modulate the input power interferes with continuous-wave
operation of an accelerator, but only very briefly, because the time needed to
achieve reasonable accuracies is very short, only a few seconds or minutes are
needed. This could either be scheduled during dedicated machine development
shifts or during brief interruptions of regular operations.
\subsection*{Acknowledgments}
Discussions with Tor Lofnes, Uppsala University are gratefully acknowledged.
\bibliographystyle{plain}

\begin{thebibliography}{M}
%
\bibitem{CEBAF}
  B. Norum, J. McCarthy, R. York, {\em CEBAF - a high-energy, high duty factor
    electron accelerator for nuclear physics,} Nucl. Instrum. Methods B10/11 (1985) 337.
\bibitem{CBETA}
    A. Bartnik, N. Banerjee, D. Burke, J. Crittenden, K. Deitrick, J. Dobbins, et al.,
  {\em CBETA: First Multipass Superconducting Linear Accelerator with
    Energy Recovery,} Phys. Rev. Lett. 125 (2020) 044803.
\bibitem{LCLS2}
  S. Posen, A. Cravatta, M. Checchin, S. Aderhold,  C. Adolphsen,  T. Arkan et al.,
  {\em High gradient performance and quench behavior of a verification
    cryomodule for a high energy continuous wave linear accelerator,}
  Physical Review Accelerator and Beams 25 (2022) 042001.
\bibitem{POWERS}
  T. Powers, {\em Theory and Practice of Cavity RF Test Systems,} Proceedings of the
  12th International Workshop on RF Superconductivity at Cornell University, Ithaka, 2005,
  SUP02, page 40.
\bibitem{MELNYCHUK}O. Melnychuk, A. Grasselino, A. Romanenko, {\em Error analysis for intrinsic
    quality factor measurements in superconducting radio frequency resonators,}
  Review of Scientific Instruments 85 (2014) 124705.
\bibitem{HOLZ2016}
  J. Holzbauer, Y. Pischalnikov, D. Sergatskov, W. Schappert, S. Smith, {\em Systematic
    uncertainties in RF-based measurement of superconducting cavity quality factors,}
  Nuclear Instruments and Methods in Physics Research A 830 (2016) 22.
\bibitem{HOLZ2019}
  J. Holzbauer, C. Contreras, Y. Pischalnikov, D. Sergatskov, W. Schappert, {\em Improved
    RF measurements of SRF cavity quality factors,} Nuclear Inst. and Methods in Physics
  Research, A 913 (2019) 7.
\bibitem{FROLOV}
  D. Frolov, {\em Intrinsic quality factor extraction of multi-port cavity with
    arbitrary coupling,} Rev. Sci. Instrum. 92 (2021) 014704.
\bibitem{GORYASHKO}
  V. Goryashko, A. Bhattacharyya, H. Li, D. Dancila, R. Ruber, {\em A method for
    high-precision characterization of the $Q$-slope of superconducting cavities,}
  IEEE Transactions on microwave theory and techniques 64 (2016) 3764.
\bibitem{KUZIKOV}
  S. Kuzikov, P. Avrakov, C. Jing, R. Kostin, Y. Zhao et al., {\em A method for in-situ Q0
    measurements of high-quality SRF resonators,} Proceedings of the 20th Conference on
  RF Superconductivity in East Lansing, 2021, page 221.
\bibitem{DRURY}
  M. Drury, T. Lee, J. Marshall, J. Preble, Q. Saulter, W. Schneider, M. Spata, M. Wiseman,
  {\em Commissioning of the CEBAF Cryomodules,} Proceedings of the Particle Accelerator Conference
  in Washington, DC, page 841.
\bibitem{WANG}
  R. Wang, B. Hansen, M. White, J. Hurd, O. Al Atassi et al., {\em Operational Experience
    from LCLS-II Cryomodule Testing,} IOP Conf.Ser.Mater.Sci.Eng. 278 (2017) 1, 012187;
  DOI:\url{10.1088/1757-899X/278/1/012187}.
\bibitem{SYSID}
  V. Ziemann, {\em Simulations of real-time system identification of superconducting cavities
    with a recursive least-squares algorithm,} Physical Review Accelerators and Beams 26 (2023)
  112003.
\bibitem{LAIWEI}
  T. Lai, C. Wei, {\em Least squares estimates in stochastic regression models with
    applications to identification and control of dynamic systems,} The Annals of
  Statistics 10 (1982) 143.
\bibitem{SCHILCHER}
  T. Schilcher, {\em Vector sum control of pulsed accelerating fields in lorentz
    force detuned superconducting cavities,} Dissertation, Universit\"at Hamburg, 1998.
\bibitem{PENROSE}
  R. Penrose, {\em A generalized inverse for matrices,} Mathematical Proceedings of the Cambridge
  Philosophical Society {\bf 51} (1955) 406 (\url{https://doi.org/10.1017/S0305004100030401}).
\bibitem{NR}
  W. Press et al., {\em Numerical Recipes, 2nd ed.}, Cambridge University Press, Cambridge, 1992.
\bibitem{MATLAB}
  Mathworks web site at \url{www.mathworks.com}
\bibitem{GITHUB}
  Github repository for the software accompanying this report: \url{https://github.com/volkziem/SysidRFcavity}.
\bibitem{AN1364-1}
  Agilent Technologies, {\em De-embedding and Embedding S-parameters Networks Using a
    Vector Network Analyzer,} Agilent Application Note 1364-1, 2004.
%
\end{thebibliography}

%
\end{document}